\documentclass[11pt,a4paper]{article}

\usepackage[hyperref]{emnlp2018}
\usepackage{times}
\usepackage{latexsym}
\usepackage{url}
\usepackage{textcomp}

\usepackage{amsmath}
\usepackage{amssymb}
\usepackage{graphicx}
\usepackage{enumitem}
\usepackage{subfig}
\usepackage{floatrow}
\usepackage[skip=0pt]{caption}
\usepackage[ruled,vlined]{algorithm2e}
\usepackage[font=small,tableposition=top]{caption}
\usepackage{colortbl}

\DeclareCaptionLabelFormat{andtable}{\tablename~\thetable \ \&  #1~#2}
\DeclareCaptionLabelFormat{andfigure}{#1~#2 \& \tablename~\thetable}

\newcommand{\nltable}{
    \begin{tabular}{l l l}
                        & \textbf{Top Words} & \textbf{Top Subreddits} \\ \hline
    Massachusetts, USA &   \begin{tabular}{@{}l@{}} allston, mbta, waltham, \\ saugus, brookline, masshole, \\ somerville, alewife, braintree  \end{tabular}  &     \begin{tabular}{@{}l@{}} \textit{r/PokemonGoBoston}, \textit{r/WorcesterMA}, \\ \textit{r/massachusetts}, \textit{r/bostonhousing} \end{tabular}           \\ \hline
    Ohio, USA   &    \begin{tabular}{@{}l@{}} ohioan, westerville, cincinnatis, \\ jenis, clevelander, graeters, \\ cuyahoga, bgsu, cbus  \end{tabular}   &   \begin{tabular}{@{}l@{}} \textit{r/uCinci}, \textit{r/ColumbusSocial}, \\ \textit{r/columbusclassifieds}, \textit{r/Columbus} \end{tabular}  \\ \hline
     Germany   &    \begin{tabular}{@{}l@{}} zeigen, dennoch, wenige, \\ zeigt, solltest, deutlich, \\ wollt, kriegt, stck  \end{tabular}   &   \begin{tabular}{@{}l@{}} \textit{r/FragReddit}, \textit{r/de\_IAmA}, \\ \textit{r/rocketbeans}, \textit{r/kreiswichs} \end{tabular}  \\ \hline
    Belgium    &    \begin{tabular}{@{}l@{}} telenet, walloon, vlaams, \\ jupiler, leuven, vlaanderen, \\ ghent, molenbeek, azerty  \end{tabular}    &  \begin{tabular}{@{}l@{}} \textit{r/belgium}, \textit{r/brussels}, \\ \textit{r/Vivillon}, \textit{r/ecr\_eu} \end{tabular}      
    
    \end{tabular}
}

\newcommand{\transferresultsvtwo}{
\begin{tabular}{l l |c c c |c c c |c c c |c c c |c c c |c c c}
                    &       \multicolumn{1}{r|}{\textbf{Test}}                         & \multicolumn{3}{c|}{\textsc{reddit-US}} & \multicolumn{3}{c|}{\textsc{reddit-Global}} & \multicolumn{3}{c|}{\textsc{Geotext}} & \multicolumn{3}{c|}{\textsc{Twitter-US}} & \multicolumn{3}{c|}{\textsc{Twitter-World} (US)} & \multicolumn{3}{c}{\textsc{Twitter-World} (All)}  \\
\textbf{Train}                    &                               & \multicolumn{1}{c}{\textit{Acc@100}} & \multicolumn{1}{c}{\textit{AED}} & \multicolumn{1}{c|}{\textit{MED}}     & \multicolumn{1}{c}{\textit{Acc@100}} & \multicolumn{1}{c}{\textit{AED}} & \multicolumn{1}{c|}{\textit{MED}}          & \multicolumn{1}{c}{\textit{Acc@100}} & \multicolumn{1}{c}{\textit{AED}} & \multicolumn{1}{c|}{\textit{MED}}                 & \multicolumn{1}{c}{\textit{Acc@100}} & \multicolumn{1}{c}{\textit{AED}} & \multicolumn{1}{c|}{\textit{MED}}                   & \multicolumn{1}{c}{\textit{Acc@100}} & \multicolumn{1}{c}{\textit{AED}} & \multicolumn{1}{c|}{\textit{MED}}         & \multicolumn{1}{c}{\textit{Acc@100}} & \multicolumn{1}{c}{\textit{AED}} & \multicolumn{1}{c}{\textit{MED}}            \\ 
\hline
\multicolumn{2}{l|}{Rahimi et al. \shortcite{rahimi2017neural} (Text-based)}  & -   & -  & -         &  -  & - & -            & 0.38   & 844  & 389          & 0.54   & 554 & 120            & -   & - & -                    & 0.34   & 1456 & 415               \\ 
\multicolumn{2}{l|}{Do et al. \shortcite{do2017multiview} (Text + Network)}  & -   & -  & -         &  -  & - & -            & 0.62   & 532  & 32          & 0.66   & 433 & 45            & -   & - & -                    & 0.53   & 1044 & 118               \\ 
\hline
\multicolumn{2}{l|}{Baseline (MAP Estimate)}                       & 0.05   & 894  & 750          &  0.03   & 2207 & 1221            & 0.33   & 679  & 424          & 0.07   & 1659 & 1869            & 0.07   & 1651 & 1869                    & 0.04   & 3893 & 2463               \\ 
\hline
\textsc{reddit-US}             & $\vec{w}$                         &  0.36   & 602  & 295          & -   & -  & -             & 0.21   & 695  & 479                     & 0.31   & 609  & 362                &  0.14   & 748  & 605            & -      & -    & -                 \\
                    & $\vec{w} + \vec{\tau}$             & 0.36   & 580  & 278          & -   & -  & -             & 0.21   & 695  & 480                     & 0.31   & 603  & 358                 & 0.13   & \textbf{747}  & 592            & -      & -    & -                 \\ 
                    & $\vec{w} + \vec{\tau} + \vec{s}$ & \textbf{0.45}   & \textbf{502}  & \textbf{157}        & -      & -    & -               & -      & -    & -                       & -      & -    & -                        & -      & -    & -              & -   & - & -              \\
                    
\hline
\textsc{reddit-Global}          &  $\vec{w}$                          & -   & -  & -          & 0.24   & 1751 & 590              & -   & -  & -                     & -      & -    & -                        & -   & -  & -            &  0.09   & 2717 & 1329                \\
                    & $\vec{w} + \vec{\tau}$                  & -   & -  & -          & 0.25   & 1475 & 457              & -   & -  & -                     & -      & -    & -                        & -   & -  & -            &  0.09   & 2708 & \textbf{1329}             \\ 
                    & $\vec{w} + \vec{\tau} + \vec{s}$   & -   & -    & -        & \textbf{0.36}   & \textbf{1259} & \textbf{266}             & -      & -    & -                & -      & -    & -                        & -      & -    & -       & -   & - & -              \\
\hline
\textsc{Geotext}  &  $\vec{w}$                          & 0.07   & 1210 & 1019           & -      & -    & -             & 0.38   & 591  & 280          & 0.12   & 982  & 755             & 0.13   & 992  & 755                     & -      & -    & -                  \\
                    & $\vec{w} + \vec{\tau}$  & 0.07   & 1209 & 1019           & -      & -    & -            & \textbf{0.38}   & \textbf{575}  & \textbf{271}          & 0.12   & 982  & 755             & 0.13   & 992  & 755                     & -      & -    & -                  \\ 
\hline
\textsc{Twitter-US} &  $\vec{w}$                          & 0.36   & 670  & 326            & -      & -    & -               & 0.34   & 631  & 301          & 0.40   & 547  & 225             & 0.24   & 790  & 583                     & -      & -    & -              \\
                    & $\vec{w} + \vec{\tau}$     & 0.36   & 635  & 294            & -      & -    & -                & 0.33   & 632  & 304          & \textbf{0.40}   & \textbf{536}  & \textbf{220}             & \textbf{0.24}   & 789  & \textbf{582}                     & -      & -    & -                \\ 
\hline
\textsc{Twitter-World} (US)           &  $\vec{w}$      & 0.20   & 963  & 729          & -   & -  & -             & 0.12   & 635  & 313          & 0.23   & 772  & 564             & 0.22   & 795  & 589                     & -      & -    & -                  \\
                    & $\vec{w} + \vec{\tau}$        & 0.18   & 923  & 717          & -   & -  & -             & 0.13   & 628  & 311          & 0.22   & 768  & 563             & 0.22   & 791  & 584                     & -      & -    & -                   \\
\hline
\textsc{Twitter-World} (All)        &  $\vec{w}$                & -      & -    & -            &  0.15   & 2737 & 1829                & -      & -    & -                       & -   & - & -                     & -      & -    & -              & 0.16   & 2716 & 1665               \\
                    & $\vec{w} + \vec{\tau}$              & -      & -    & -              & 0.16   & 1793 & 817              & -      & -    & -                       & -   & - & -                     & -      & -    & -              & \textbf{0.16}   & \textbf{2610} & 1405             
\end{tabular}
}

\newcommand{\countrybreakdown}{
\begin{tabular}{c c c}
\textbf{Country}     & \textbf{Alexa Traffic} & \textbf{Labeled Users} \\ \hline
United States  & 58.7\%     & 60.1\% (n=39,236)     \\
United Kingdom & 7.4\%      & 5.4\%    (n=3,544)     \\
Canada         & 6.0\%      & 9.4\%  (n=6,163)    \\
Australia      & 3.1\%      & 3.5\% (n=2,344)       \\
Germany        & 2.1\%      & 1.7\%   (n=1,097)   
\end{tabular}
}

\aclfinalcopy 

\title{Geocoding Without Geotags: A Text-based Approach for reddit}

\author{Keith Harrigian \\
  Warner Media Applied Analytics\\
   Boston, MA \\
  {\tt keith.harrigian@appliedanalytics.net}
  \\}

\date{}

\begin{document}

\maketitle


\setlength{\textfloatsep}{4pt plus 1.0pt minus 2.0pt}

\setlength{\abovedisplayskip}{1pt}
\setlength{\belowdisplayskip}{1pt}
\setlength{\abovecaptionskip}{-2pt}
\setlength{\belowcaptionskip}{-2pt}
\renewcommand{\arraystretch}{1.25}


\begin{abstract}

In this paper, we introduce the first geolocation inference approach for reddit, a social media platform where user pseudonymity has thus far made supervised demographic inference difficult to implement and validate. In particular, we design a text-based heuristic schema to generate ground truth location labels for reddit users in the absence of explicitly geotagged data. After evaluating the accuracy of our labeling procedure, we train and test several geolocation inference models across our reddit data set and three benchmark Twitter geolocation data sets. Ultimately, we show that geolocation models trained and applied on the same domain substantially outperform models attempting to transfer training data across domains, even more so on reddit where platform-specific interest-group metadata can be used to improve inferences. 

\end{abstract}


\section{Introduction}

The rise of social media over the past decade has brought with it the capability to positively influence and deeply understand demographic groups at a scale unachievable within a controlled lab environment. For example, despite only having access to sparse demographic metadata, social media researchers have successfully engineered systems to promote targeted responses to public health issues \cite{yamaguchi2014online, huang2017examining} and to characterize complex human behaviors \cite{mellon2017twitter}. 

However, recent studies have demonstrated that social media data sets often contain strong population biases, especially those which are filtered down to users who have opted to share sensitive attributes such as name, age, and location \cite{malik2015population, sloan2015tweets, lippincott2018observational}. These existing biases are likely to be compounded by new data privacy legislation that will require more thorough informed consent processes \cite{kho2009written, gdprOfficial}. 

While some social platforms have previously approached the challenge of balancing data access and privacy by offering users the ability to share and explicitly control public access to sensitive attributes, others have opted not to collect sensitive attribute data altogether. The social news website reddit is perhaps the largest platform in the latter group; as of January 2018, it was the 5th most visited website in the United States and 6th most visited website globally \cite{alexa2018}. Unlike real-name social media platforms such as Facebook, reddit operates as a pseudonymous website, with the only requirement for participation being a screen name.

Fortunately, there has been significant progress made using statistical models to infer user demographics based on text and other features when self-attribution data is sparse \cite{han2014text, ajao2015survey}. On reddit in particular, Harrigian et al. \shortcite{harrigian_gender} has used self-attributed ``flair" as labels for training a text-based gender inference model. However, to the best of our knowledge, user geolocation inference has not yet been attempted on reddit, where a complete lack of location-based features (e.g. geotags, profiles with a location field) has made it difficult to train and validate a supervised model. 

The ability to geolocate users on reddit has substantial implications for conversation mining and high-level property modeling, especially since pseudonymity tends to encourage disinhibition \cite{gagnon2013disinhibition}. For instance, geolocation could be used to segment users discussing a movie trailer into US and international audiences to estimate a film's global appeal or to inform advertising strategy within different global markets. Alternatively, user geolocation may be used in conjunction with sentiment analysis of political discussions to predict future voting outcomes. 

Moving toward these goals, we introduce a text-based heuristic schema to generate ground truth location labels for reddit users in the absence of explicitly geotagged data. After evaluating the accuracy of our labeling procedure, we train and test several geolocation inference models across our reddit data set and three benchmark Twitter geolocation data sets. Ultimately, we show that geolocation models trained and applied on the same domain substantially outperform models attempting to transfer training data across domains, even more so on reddit where platform-specific interest-group metadata can be used to improve inferences. 


\section{reddit: the front page of the internet} \label{redditbackground}

Founded originally as a social news platform in 2005, reddit has since become one of the most visited websites in the world, offering an expansive suite of features designed to ``[bridge] communities and individuals with ideas, the latest digital trends, and breaking news" \cite{redditGeneral}. Although the so-called \textit{front page of the internet} still boasts an impressive amount of link-sharing, reddit has gradually evolved into a self-referential community where original thoughts and new content prevail \cite{singer2014evolution}. 

reddit is structured much like a traditional online forum, where over one-hundred thousand topical categories known as subreddits separate user communities and conversation. Subreddits may cover topics as general as humor and movies (e.g. \textit{r/funny}, \textit{r/movies}) or as specific as an unusual fitness goal (e.g. \textit{r/100pushups}). Within each subreddit, users can post thematically relevant submissions in the form of an image, text blurb, or external link.  Users are then able to post comments on the submission, responding to either the content in the original submission or to comments made by other users.

reddit users tend to feel protected by the site's pseudonymity and consequently eschew their offline persona in favor of a more genuine online persona \cite{gagnon2013disinhibition, shelton2015online}. Thus, there is much potential in reddit as a data source for conversation mining. Unfortunately, the same policies which promote this favorable behavior currently preclude the segmentation of conversation based on demographic dimensions and serve as a fundamental motivation to our research.


\section{Geocoding reddit Users}

Previous research on geolocation inference for social media has primarily used three Twitter data sets for model training and validation \cite{eisenstein2010latent, roller2012supervised, bo2012geolocation}. Although Twitter's topical diversity typically supports generalization to other platforms \cite{mejova2012crossing}, it has not been tested thoroughly in the geolocation context or on reddit at all. 

There is substantial reason to believe that geolocation models trained on Twitter data will not perform optimally when applied to reddit. One of the most glaring concerns is the variation in user demographics between the platforms. Namely, Twitter tends to skew female while reddit tends to skew male \cite{pewRedditUsers, pew2018social}. Coates and Pichler \shortcite{coates2011language} have shown that language varies between genders, which suggests that language-model priors may vary based on social platform.

Additionally, the lack of geolocation ground truth for reddit necessarily excludes several promising inference models from being applied to the platform. For example, network-based geolocation approaches exploit the empirical relationship between physical user distance and connectedness on a social graph to propagate known user locations to unlabeled users \cite{backstrom2010find}. Rahimi et al. \shortcite{rahimi2015exploiting} argue that network-based approaches are generally superior to content-based methods, especially when the social graph is well-connected. However, these models require within-domain grounding for the propagation algorithms to be useful. 

Finally, while the reddit platform lacks certain features useful for geolocation on Twitter (e.g. timezones, profile location fields), it possesses its own unique assets which we hypothesize can prove useful for user geolocation. In this paper, we quantify the predictive value of subreddit metadata, leaving other features such as user flair and the platform's hierarchical comment structure for future research.

\subsection{Labeling Procedure} \label{labeling}

\begin{figure}[t]
\includegraphics[width=\columnwidth]{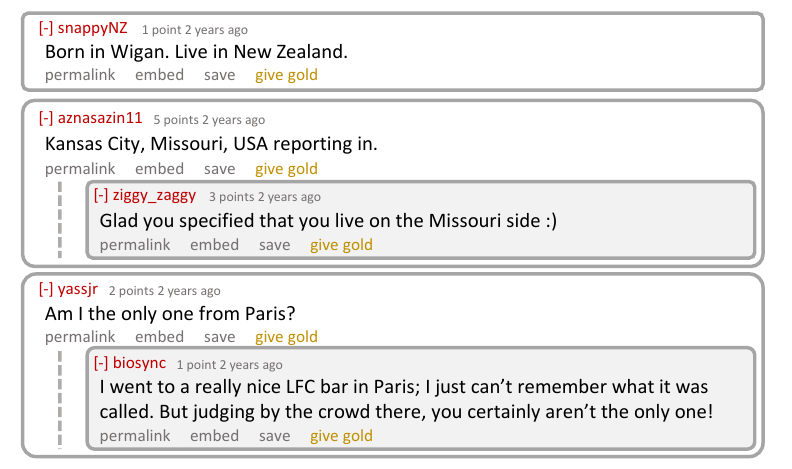}
\vspace{-16pt}
\caption{Comments from one of the seed submissions. Replies (gray background) are filtered out because they often contain location names not representative of user location.}
\label{example_thread}
\end{figure}

Since reddit does not offer geotagging capabilities, nor do reddit user profiles include location fields, we design a free-text geocoding procedure to associate reddit users with a home location. While we focus on reddit for this paper, we believe that the following labeling method should generalize to other pseudonymous platforms that have a question and answer-based submission structure.

\textbf{Data.} We begin by querying the reddit API for submissions with a title similar to ``Where are you from?", ``Where do you live?", or ``Where are you living?" This query yields 3,600 English-language submissions, from which we manually curate a subset of 1,200 seed submissions where we expect users to self-identify their home location.  Examples of submissions within the final set include ``Where do you support Liverpool from?" and ``How much is your rent, and where do you live?" Examples of submissions in the original query which were excluded from the final set include ``Where are you banned from?" and ``Without naming the location, where are you from?"

We then query comment data from the 1,200 seed submissions using the Python Reddit API Wrapper (PRAW)\footnote{https://praw.readthedocs.io/en/latest/}. We filter out comments which mention ``move," ``moving," ``born," or ``raised" to ensure only current home locations are identified. We also remove comment replies, finding that they often include location mentions from a perspective of discussion as opposed to self-identification. Representative comments from one of the seed submissions\footnote{www.reddit.com/r/LiverpoolFC/comments/3wmjd4/} are displayed in Figure 1. Ultimately, we keep 96,071 comments from 89,697 unique users.

\textbf{Entity Extraction.} We employ a string matching approach to identify locations mentioned within the comment text. As a gazetteer, we use a subset of the \textsc{Geonames} data set \cite{wick2012geonames} with city populations greater than 15,000. To reduce the false positive rate of location recognition, 90 location names found within the 5,000 most common English words from the \textsc{Corpus of Contemporary English} \cite{davies2009385+} are removed from the gazetteer. After applying this filter, our gazetteer is left with 23,018 cities and their associated location hierarchies (i.e. city, state, country). 

To aid in the disambiguation of common location names, we create a small dictionary of 57 frequently occurring abbreviations found during a preliminary exploration of the data. This dictionary includes abbreviations for each state and territory of the United States, in addition to the following: USA (United States), UK (United Kingdom), BC (British Columbia), and OT (Ontario). Abbreviations are only extracted if they directly follow an $n$-gram found within the location gazetteer.

Each comment is tokenized into an ordered list using standard processing techniques---contraction and case normalization, number and hyperlink removal, and whitespace splitting. To support identification of multi-token locations, we then chunk each list into all possible $n$-grams for $n \in [1,4]$. We remove any $n$-gram which does not have an exact match to a location in the gazetteer or the abbreviation dictionary. When two or more of the matched $n$-grams occur in an order of a known location hierarchy from \textsc{Geonames}, they are concatenated together. Any remaining $n$-gram which is a substring of a larger $n$-gram in the list of matches is removed. 

\begin{figure*}[!ht]
\centering
\raisebox{-0.7\height}{\includegraphics[width=.45\columnwidth]{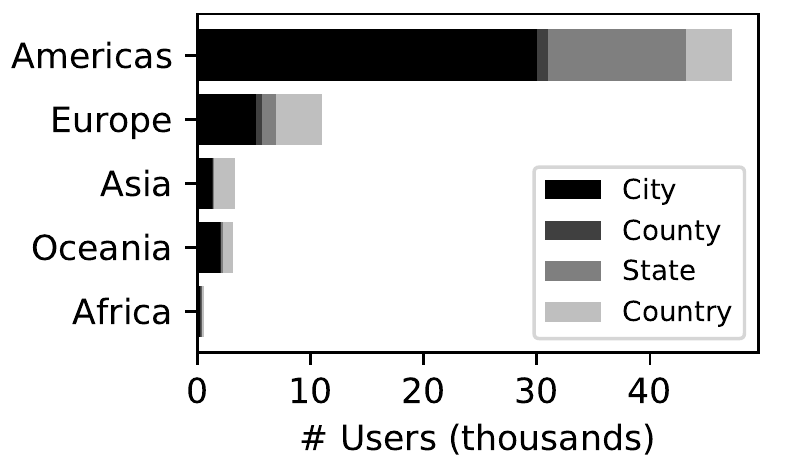}}
\qquad
\raisebox{-0.3\height}{\resizebox{.49\columnwidth}{!}{\countrybreakdown}} 
\captionlistentry[table]{A table beside a figure}
\captionsetup{labelformat=andfigure}
\caption{The figure (left) shows the geographic distribution of labeled users. Users outside the Americas are generally more likely to self-identify using a state-level resolution and above. The table (right) compares relative reddit traffic estimated using the proprietary Alexa \shortcite{alexa2018} panel to the distribution within our labeled data set.}
\label{table:label_dist_figure}
\vspace{-6pt}
\end{figure*}

One issue with this approach is that cities under the population threshold, or cities missing from our gazetteer, are ignored entirely. However, we note that users often reference their home location in the form \textit{City, State, Country}. To improve our labeling procedure's recall, we devise an additional rule that captures location mentions which match this syntactic pattern and contain at least one of our gazetteer or abbreviation dictionary entries as a substring. The inclusion of this rule expands the set of users with estimated city-level locations from 25k to 38k.

\textbf{Geocoding.} To associate each $n$-gram with a coordinate position, we employ Google's Geocoding API, which ingests a string and returns the most probable coordinate pair and nominal location hierarchy information. To help disambiguate common location names, we manually map region biases to a subset of location-related subreddits that house the seed submissions. We feed these region biases into the Geocoding API as a query parameter for comments which were made in any of the mapped subreddits. Thus, a comment that mentions Scarborough in \textit{r/ontario} will be properly mapped to Canada, while a comment that mentions Scarborough in \textit{r/CasualUK}, will be mapped to England.

In the final stage of our procedure, we aggregate location extractions across each user, taking the maximum overlap within their identified location hierarchies as the ground truth. For each location above a city-level resolution, we use the geodesic median \cite{vardi2000multivariate} of labeled users at the given resolution and below as the true coordinate pair.

\subsection{Geographic Representation}

Ultimately, we associate 65,245 users with geolocation labels at varying maximum resolutions---38,773 at a city level; 1,541 at a county level; 13,774 at a state level; and 11,157 at a country level. The label distribution and resolution breakdown over continents can be seen in Figure 2. 

While the underlying geographic distribution of reddit users is not publicly available, Alexa \shortcite{alexa2018} publishes an estimate of country-level activity for the top-5 sources of reddit traffic using a proprietary data panel. We find that the same countries make up the top-5 most represented nations in our labeled data set, albeit having slightly different proportional breakdowns (see Table 1). In particular, we note that our data set slightly over-indexes on North American reddit users and suffers from a low sample size for most countries outside of the top-5. 



\subsection{Label Accuracy}

To estimate precision of our labeling procedure across the larger data set, we randomly sample 500 of the labeled users and the comment from which their labeled location was extracted. For each of the 500 comments, we hand label whether the user's home location was properly extracted and, if correct, whether the location was extracted at the appropriate resolution.

Formally, we score any extracted location which falls within the true location hierarchy as a correct label. For example, if the true location is Boston, Massachusetts, but our procedure extracts Massachusetts (the state), we score the label as correct, but at an incorrect resolution. We find that our labeling procedure correctly extracted and geocoded 96.6\% of the location names from the raw comments. Of these correct labels, 92.55\% were labeled at the correct resolution.

Of the false positives in our random sample, 8 were due to disambiguation errors on part of the Google Geocoding API, 7 were due to users referencing locations that were not their actual home location, and 2 were due to usage of a common word not filtered out using the \textsc{Corpus of Contemporary English}. Future work will look into ways to effectively leverage more granular subreddit-location mappings during the geocoding procedure to aid in disambiguation. This exploration may prove most helpful for ambiguous locations which are housed within the same national region (e.g. Kansas City, MO and Kansas City, KS).

For labels extracted within the appropriate hierarchy, but at the incorrect resolution, there were two main sources of error. First, over half of the incorrect resolution labels were due to our system (as designed) aggregating multiple location mentions within a comment up to the maximum overlap present. Second, several of the true cities did not exist within the gazetteer and were not captured by our syntactic rule. More extensive named entity resolution systems may be useful in addressing the issue of missing gazetteer entries. However, correctly handling multiple location names and location names not representative of a user's home location will likely require a more robust natural language understanding system.
 

\section{Geolocation Inference} 

In the remainder of the paper, we evaluate whether our ``imperfect" geolocation labels are still useful in the context of a common task---user geolocation inference. We begin this analysis with a series of experiments and qualitative diagnostics to understand geolocation inference performance when training and applying models within the reddit domain. To quantify the effect of domain transfer, we conclude with a comprehensive comparison of inference performance across three benchmark Twitter data sets and our new reddit data set.

\subsection{Related Work}

Accounting for variations in feature selection and model architecture, existing geolocation inference approaches broadly fall into the following three categories (and their hybridizations): network-based, content-based, and metadata-based. We refer the reader to Han et al. \shortcite{han2014text} and Ajao et al. \shortcite {ajao2015survey} for a comprehensive overview of the literature, but will highlight research pertinent to this particular study below. We ignore network-based models because they are not relevant to our chosen inference architecture.

\textbf{Content-based.} Content-based approaches, in which geographically predictive features are extracted from user-generated multimedia and modeled thereafter, remain some of the most fundamental to user geolocation on social media. Drawing upon well-documented phenomena regarding lexicon usage and its geographical variation \cite{trudgill1974linguistic, vaux2003harvard}, text-driven models are particularly suited for application on reddit, where written comments are the lowest level of user behavior data accessible via the platform's public API.

One of the earliest contributors to user geolocation inference for social media, Cheng et al. \shortcite{cheng2010you} introduced a generative model that operates on word usage alone to infer a user's home location, achieving an average prediction error of 535 miles for US Twitter users. Chang et al. \shortcite{chang2012phillies} extended this work by replacing frequency-based word likelihoods with smoothed estimations using Gaussian Mixture Models (GMM). We use this approach within our evaluation due to its ease of implementation and interpretability. However, others have substantially improved inference accuracy using more flexible modeling architectures such as spatial topic models \cite{eisenstein2010latent, hu2013spatial}, stacked denoising autoencoders \cite{liu2015estimating}, sparse coding and dictionary learning \cite{cha2015twitter}, and most recently, neural networks \cite{rahimi2017neural}.

\textbf{Metadata-based.} We define metadata as all user behavior not explicitly expressed in multimedia, such as text or image posts, nor directly encoded as a social network connection. While metadata has generally been used to improve performance of content- and network-based methods, there exist some cases where metadata-based models outperform competing approaches outright \cite{han2014text, dredze2016geolocation}. With reddit only recently introducing user profiles to the platform and their adoption by the larger community still an open question \cite{shelton2015online}, we focus primarily on comment metadata.

First, we suspect that the subreddits a user posts in, which often cover geographically localized topics such as sports and news, will be predictive of user location. While subreddits are specific to the reddit platform, they theoretically align with group membership, which has been shown to correlate with and predict user location in other domains \cite{zheleva2009join, chen2013interest}. 

Of additional interest is temporal metadata, which can capture longitudinal variations in cyclical user activity patterns \cite{gao2013exploring} or identify geographically-centered and time-dependent events \cite{yamaguchi2014online}. In particular, Dredze et al. \shortcite{dredze2016geolocation} and Do et al. \shortcite{do2017multiview} find self-identified timezone information to be a useful geolocation predictor. Unfortunately, this explicitly encoded geographic indicator is absent from reddit and thus motivates our work to transform and model raw timestamp data in Section \ref{estimation_design}.

\subsection{Location Estimation Model} \label{estimation_design}

Each user is represented by a concatenation of three feature vectors: word usage $\vec{w}$, subreddit submissions $\vec{s}$, and posting frequency $\vec{\tau}$. $\vec{w}$ is constructed by concatenating all comment text for a user, tokenizing into uni-grams (using the same process in \ref{labeling}), and counting token frequencies. $\vec{s}$ is simply the frequency distribution of subreddits that a user has posted in across their comment history. 

We represent the temporal posting habits of a user by $\vec{\tau}$, a 24-dimensional vector where each index contains the comment counts for one hour of the day. reddit comment timestamps are reported in Coordinated Universal Time (UTC) and can therefore be interpreted uniformly across users when constructing $\vec{\tau}$. Moving forward, we let $\vec{u}$ represent the concatenation of $\vec{w}$ and $\vec{s}$, allowing either modality to be turned off within the model. However, we keep $\vec{\tau}$ separate for notational clarity.

\textbf{Model Architecture.} As mentioned briefly above, we use the generative model introduced by Cheng et al. \shortcite{cheng2010you} as the basis for our work, modifying it to enable ingestion of temporal metadata. Formally, given a user with feature set $\vec{u}$ (each $u$ having an occurrence count $\| u \|$) and posting frequency $\vec{\tau}$, we estimate the probability of the user being located at geographic coordinate pair $c$ using the following model:

\begin{small}
\begin{equation}
P(c|\vec{u}, \vec{\tau}) \propto P(c|  \vec{\tau}) \sum\limits_{u \in \vec{u}}  \| u \| P(c|u)P(u).
\end{equation}
\end{small}

\noindent To make a singular location prediction, we take the $argmax$ of $P(c|\vec{u}, \vec{\tau})$ over an arbitrarily discrete set of coordinate pairs $C$.

\textbf{Probability Density Estimation.} As a baseline, Cheng et al. \shortcite{cheng2010you} use the count-based frequency of features over cities in their training data to estimate $P(c|u)$. They recognize as a shortcoming to this approach the issue of feature sparsity---features with a low frequency or selective location presence contribute a probability close to zero for several candidate locations. 

Drawing upon the relationship of geographic query dispersion quantified by Backstrom \shortcite{backstrom2008spatial}, Chang et al. \shortcite{chang2012phillies} and Priedhorsky et al. \shortcite{priedhorsky2014inferring} use a bivariate Gaussian Mixture Model (GMM) to estimate $P(c|u)$ and demonstrate a significant performance improvement over the Cheng et al. \shortcite{cheng2010you} baseline approach. However, the authors highlight as a potential caveat to their work the sensitivity of performance to GMM hyper-parameters, namely the number of mixture components.

To mitigate this issue, we use a Dirichlet Process Mixture Model (DPMM) to estimate $P(c|u)$. Generally, DPMM is able to better describe mixtures with a varying number of components by constructing an infinite mixture with some components damped to near-zero amplitude; additionally, learned DPMM parameters remain relatively stable regardless of hyper-parameter choice \cite{attias2000variational, blei2006variational}. 

We perform a series of experiments to compare performance between GMM and DPMM. Holding constant the hyper-parameter for number of mixture components\footnote{For DPMM, we use a truncated distribution with a maximum number of mixture components equal to the chosen number of mixture components for GMM.}, we find DPMM reduces error relative to GMM on a fixed test set by 5\% to 32\% depending on the type of covariance matrix used (i.e. diagonal, spherical). Ultimately, we use DPMM with a diagonal covariance matrix and 5 components to optimize inference performance. 


\begin{figure}[t]
\includegraphics[width=\columnwidth]{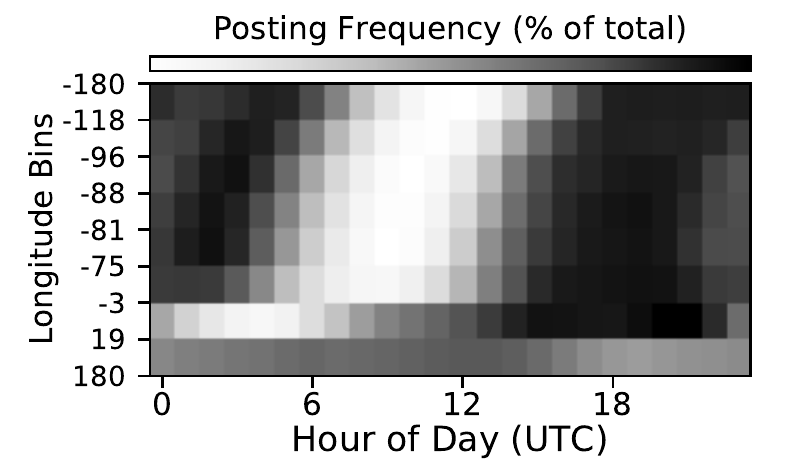}
\vspace{-16pt}
\caption{The relative posting frequencies within each longitude bin $\ell \in L$. Users east of 3\textdegree\ are more likely to post between the 6th and 12th hours (UTC) of the day.}
\label{time_dist}
\end{figure}

\textbf{Temporal Variation}. A significant change to the Chang et al. \shortcite{cheng2010you} model is the addition of a temporal adjustment term $P(c|\vec{\tau})$, designed to re-weight the word and subreddit posterior according to how well a user's posting frequency $\vec{\tau}$ aligns with the posting frequency distribution for users at coordinate pair $c$.

To make this estimate, we begin by discretizing the longitudes of users within our training data into a set of percentile-based bins $L$. Then, we fit a Logistic Regression model to map each user's posting frequency vector $\vec{\tau}$ to their associated longitude bin $\ell \in L$, using cross-validation to select hyper-parameters (e.g. L2-regularization) that maximize classification accuracy.  Additionally, we use 5-fold cross-validation to select the number of longitude bins $\| L \|$ that minimizes downstream inference error. 

To use the trained temporal feature model within our estimator, we first estimate $P(\ell | \vec{\tau})$ for each $\ell \in L$.  Then we assign each $c \in C$ to its appropriate longitude bin $\ell$ and map the predicted probability distribution across $C$. We visualize the temporal variation across longitude bins in our reddit data set within Figure 3. 

Paralleling shifts in timezone, we note that peak user activity occurs earlier (within the UTC normalization) for users in eastern longitude bins than users in western longitude bins. Additionally, users in the eastern hemisphere are significantly more likely to post between the 6th and 12th hours (UTC) of the day.


\section{Within-Domain Evaluation}


We first examine geolocation inference performance within the domain of our reddit data set. We query all comment data for users within our set of geolocation labels using a publicly available corpus of reddit comments made between December 2005 and May 2018 \cite{redditCorpus}. For each user, we keep a maximum of 1000 comments posted up to one-month after they commented in the set of submissions used for labeling; this date-based filtering is done to mitigate the effect of users moving after posting in the seed set of submissions. Additionally, to ensure our model is not overtly biased by toponym mentions, we remove comments that were used as a part of the labeling procedure. 

We separate our reddit data set into two versions---\textsc{US} (restricted to users from the contiguous United States) and \textsc{Global} (no location restrictions). We require that users within the United States have a minimum city-level resolution, while users outside the United States have been labeled with at least a state-level resolution.

We quantify model performance using three standard metrics from the user geolocation literature: Average Error Distance (\textit{AED}), Median Error Distance (\textit{MED}), and Accuracy at 100 miles (\textit{Acc@100}). \textit{AED} and \textit{MED} are simply the arithmetic mean and median of error between predicted coordinates and true coordinates, respectively. \textit{Acc@100} is the percentage of users whose predicted location is less than 100 miles from their true location. 

\subsection{Results}

\begin{table}[t]
\resizebox{\columnwidth}{!}{\nltable}
\vspace{-4pt}
\caption{Examples of the top features ranked using non-localness. Toponyms and non-English tokens are often the most indicative of location.}
\label{time_dist}
\end{table}

\textbf{Feature Selection}. Dimensionality reduction methods have been used to improve geolocation inference performance while reducing computational cost \cite{cheng2010you, chang2012phillies, bo2012geolocation}. Of existing approaches, the non-localness ($NL$) criteria introduced by Chang et al. \shortcite{chang2012phillies} stands out as being both effective at improving inference performance and useful as a means to understand feature alignment. Formally, $NL$ is computed according to Equation \ref{eq:non_localness}, where $sim_{SKL}$ is the Symmetric Kullback-Liebler divergence, $S$ is a set of ``stopword-like" features which are expected to occur uniformly across locations, and $f$ is a generic feature in a larger feature set $F$. 

\begin{small} 
\begin{equation} \label{eq:non_localness}
NL(f) = \sum_{s \epsilon S} sim_{SKL}(f,s)\frac{count(s)}{\sum_{s' \epsilon S'}count(s')}
\end{equation}
\end{small}

We apply non-localness to $\vec{w}$ and $\vec{s}$ separately to control how many features from each modality are kept. For $\vec{w}$, we let $S$ be a set of 130 English stopwords taken from the Natural Language Toolkit\footnote{http://www.nltk.org/}; to apply non-localness to subreddit features $\vec{s}$, we assume the 30 most active subreddits\footnote{Examples include \textit{r/Politics}, \textit{r/AskReddit}, and \textit{r/funny}.} in our data set make up $S$.

We evaluate $NL$ over a discretized set of ``State, Country, Continent" combinations, rolling up locations with less than 50 users in the training data to the next level of the location hierarchy. While Chang et al. \shortcite{chang2012phillies} finds modest performance improvements using GMM to estimate feature frequencies in the $sim_{SKL}$ computation, we limit ourselves to frequency-based feature likelihoods to reduce computational expense.

In alignment with previous research, we find that $NL$ produces qualitatively intuitive feature rankings (see Table 2). Furthermore, dimensionality reduction of both words and subreddits using $NL$ significantly reduces error compared to using the full feature set for \textsc{US} and \textsc{Global}. The most significant source of error for large feature set sizes is noise added by ``stopword-like" features, which generally have a large $\| u \|$ and effectively negate the contribution of more geographically predictive features. 

Final feature set sizes are selected to minimize \textit{AED} --- 40k words and 650 subreddits for \textsc{US} (originally 118k words and 13k subreddits); 50k words and 1.1k subreddits for \textsc{Global} (originally 120k words and 14k subreddits). 


\begin{figure}[t]
\includegraphics[width=\columnwidth]{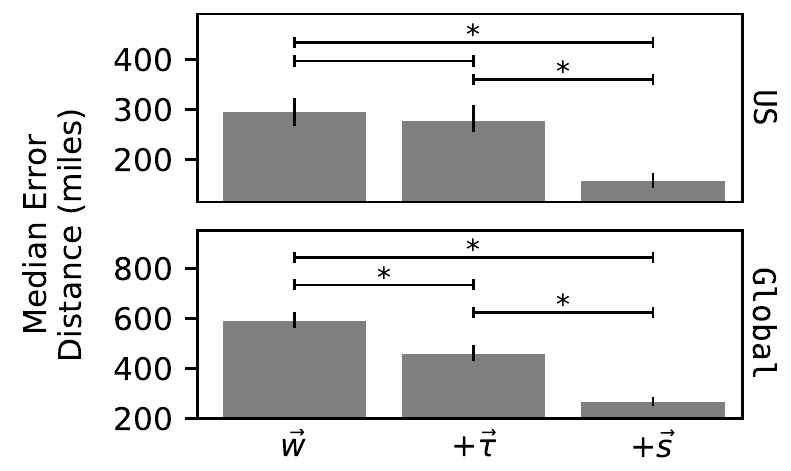}
\vspace{-16pt}
\caption{The effect of temporal and subreddit metadata on inference error.  Temporal features do not affect model performance on the \textsc{US} data set, but reduce error for the \textsc{Global} data set; subreddit metadata improves performance on both data sets.}
\label{time_dist}
\end{figure}

\textbf{Feature Modalities.} Below, we discuss how the addition of the temporal adjustment factor $P(c|\vec{\tau})$ and subreddit features $\vec{s}$ affect model performance. We carry out 5-fold cross validation, with splits varied for \textsc{US} and \textsc{Global}, but held constant within each data set, to evaluate the modality effects fairly. 

As seen in Figure 4, the temporal adjustment factor does not significantly affect model performance on the \textsc{US} data set, but reduces error when included within the \textsc{Global} data set, where underlying differences in $\vec{\tau}$ are magnified across continents. The most significant reduction in error occurs for European users, whose posting levels tend to peak around the 10th hour (UTC) of the day. 

While subreddit features reduce error within both data sets when combined with word features, they do not outperform word features on their own. Rather, models trained using word features alone achieve an \textit{AED} which is 17\% and 4\% lower for \textsc{US} and \textsc{Global}, respectively, than models trained using subreddit features alone. This implies that text-based models trained on other domains may perform adequately on reddit, but will likely suffer from the inability to take advantage of subreddit metadata in a supervised manner. 


\section{Cross-Domain Evaluation}

\begin{table*}[!t]
\centering
\resizebox{\columnwidth}{!}{%
\transferresultsvtwo
}
\caption{Summary statistics for the domain-transfer experiment. The best results from our cross-validation procedure are bolded. Models trained on reddit data (third and fourth rows) outperform the baseline for Twitter data sets in nearly all cases (see text for caveats). Within the reddit data sets (first and second columns), models with access to platform-specific metadata outperform all models transferred from the Twitter domain.}
\label{transferres}
\end{table*}

While within-domain experiments suggest that reddit-specific metadata offers substantial predictive value, we wish to compare the highest degree of performance achieved within-domain to performance achieved using models trained outside the reddit domain. To do so, we use three benchmark Twitter geolocation data sets---\textsc{Geotext} \cite{eisenstein2010latent}, \textsc{Twitter-US} \cite{roller2012supervised}, and \textsc{Twitter-World} \cite{bo2012geolocation}. 

All three data sets were created by monitoring Twitter's streaming API over a discrete period of time and caching comments for users who enabled geotagging features on the Twitter platform. Due to Twitter's terms of service, \textsc{Twitter-US} and \textsc{Twitter-World} must be compiled from scratch using tweet IDs. Unfortunately, several users within the original data sets have since either deleted their accounts or restricted access to their tweet history. As such, we were not able to perfectly recreate the original data sets. Ultimately, our compilation of \textsc{Twitter-US} contains 246k out of the original 440k users, while \textsc{Twitter-World} contains 888k out of the original 1.4M users. 

To understand the impact that domain transfer has on geolocation inference performance, we set up a systematic model comparison. First, we run 5-fold cross-validation within each of the Twitter data sets using both word and timestamp features. For the \textsc{Twitter-World} data set, we run two independent cross-validation procedures, evaluating on the subset of US users alone and also on the entire data set. Then, we train models for each of our data sets using all available data and apply them to the data sets not used for training. When evaluating performance on a data set that only contains US users, we train the corresponding model only using US user data. All model hyper-parameters (e.g. feature set sizes, regularization, etc.) were chosen to optimize within-data-set performance.

\subsection{Results}

Experimental results are summarized in Table 3 and explored in greater detail below. The baseline model assigns each user in the test set to the maximum a posteriori (MAP) of a DPMM fit to the locations of all users in the training data. As an additional reference point, we also include results from two recent approaches for the Twitter data sets.

\textbf{Transfer Performance.} In validation of our labeling procedure, we note that models trained on reddit data outperform within-domain baselines for nearly all Twitter data sets. The only exception occurs for \textsc{Geotext}, where less than 10\% of reddit features are also present. Due to the lack of feature overlap, many of the \textsc{Geotext}, users are assigned to the MAP of users in the reddit data by default. While \textsc{Twitter-US} and \textsc{Twitter-World} also have low feature overlap with \textsc{Geotext}, their MAP estimates are much closer to majority of users in \textsc{Geotext}. 

We also note that there is a significant loss incurred by most domain transfers. This loss is magnified for models trained on Twitter data and applied to reddit data, since Twitter models critically lack access to subreddit metadata during training. 

\textbf{Temporal Features.} The effect of our temporal adjustment term $P(c|\vec{\tau})$ varies between each data set. Specifically, the temporal features significantly improve within-domain performance for both of our ``international" data sets (\textsc{Twitter-World} and \textsc{reddit-Global}), but offer no significant gain for data sets with US users only. Additionally, we note that temporal features significantly reduce the loss in performance incurred by domain transfer going from \textsc{Twitter-US} to \textsc{reddit-US} and from \textsc{Twitter-World} to \textsc{reddit-Global}.

\textbf{Location Estimator.} Based on within-domain performance for each of the Twitter data sets, we recognize that our inference modeling approach is below state of the art. For example, in the space of text-only models, Rahimi et al. \shortcite{rahimi2017neural} have achieved an \textit{Acc@100} of 0.34 on \textsc{Twitter-World} using a multilayer perceptron and \textit{k-d} tree discretization over the label set. 

The performance gap between our model and state of the art approaches widens when considering multi-modal architectures. Notably,  Do et al. \shortcite{do2017multiview} have achieved an \textit{Acc@100} of 0.62 on \textsc{Geotext} using multi-view neural networks that simultaneously leverage text, profile metadata, and social network connections. Thus, we hypothesize that implementing models which are more complex than our current architecture will magnify the performance gain achieved by including subreddit metadata alongside text-based features.


\section{Discussion and Future Work}

In this paper, we introduced the first user geocoding and geolocation inference approach for reddit, demonstrating that pseudonymity is not an exhaustive barrier to supervised learning. In addition to designing a labeling procedure capable of geocoding user home locations in noisy comment data with a precision of 0.966, we demonstrated that reddit-specific metadata can be used to significantly improve inferences. Ultimately, we trained a multi-modal inference model which achieves a median error of 157 miles and 266 miles for US and international reddit users, respectively. 

Moving forward, we plan to thoroughly examine underlying biases that may exist within the users identified by our labeling procedure. Specifically, we will build on the work of Sloan and Morgan \shortcite{sloan2015tweets} and Lippincott and Carrell \shortcite{lippincott2018observational} to understand differences in user activity, interests, and conversation topicality, relative to the general reddit population. We also plan to explore different seed-submission sampling methods to improve the representation of non-North American users.


%


\bibliography{emnlp2018}
\bibliographystyle{acl_natbib_nourl}


\end{document}